# A Critical Review on Sustainable Modification of Cellulose: Why Renewability is not enough.

*Kelechukwu N. Onwukamike,*[a,b] *Stéphane Grelier,*[b] *Etienne Grau,*[b] *Henri Cramail,*[b*] *Michael A.R. Meier*[a*]

[a] Institute of Organic Chemistry (IOC), Materialwissenschaftliches Zentrum (MZE), Karlsruhe Institute of Technology (KIT), Straße am Forum 7, 76131 Karlsruhe, Germany

[b] Laboratoire de Chimie des Polymères Organiques, Université de Bordeaux, UMR5629, CNRS - Bordeaux INP - ENSCBP, 16 Avenue Pey-Berland, 33607 Pessac Cedex France

**Abstract**

As we mark the 20[th] anniversary after the introduction of the twelve principles of green chemistry, sustainable modification of cellulose, being the most abundant bio-based polymer, is worth consideration. Many researchers work on this renewable polymer, however the use of non-sustainable solvents, reactants and modification approaches simply shifts the environmental burden to other stages of the life cycle. Therefore, to achieve true sustainable modification of cellulose, its renewability combined with mild and efficient reaction protocols is crucial in order to obtain sustainable materials that will reduce the overall negative effect of the fossil-based resources they are replacing.

**Introduction**

The increasing awareness of the negative environmental effects of fossil fuels, as well as their unsustainability has over the years led to an increased interest in renewable resources. Cellulose being the most abundant bio-based polymer[1] offers a viable alternative. It constitutes 35-50% of the more than 170 x 10[9] tons of lignocellulosic biomass produced annually.[2] Cellulose is a linear homopolymer consisting of β-1,4 linked glucopyranose units.[1] It shows good mechanical and thermal properties, is equally biodegradable and biocompatible.[1] However, despite its abundance, it suffers from insolubility in common solvents, including water. Furthermore, the absence of any thermal transition makes it non-processable. This insolubility can be attributed to its inherent strong *intra-* and *inter-*molecular hydrogen bonds.[1] Thus, only solvents capable of interrupting these hydrogen bonds are capable of solubilizing cellulose. However, most of these "cellulose solvents" such as *N,N*-dimethylacetamide-lithium chloride (DMAc-LiCl),[3] dimethyl sulfoxide-tetrabutyl ammonium fluoride (DMSO-TBAF),[4] or *N*-methylmorpholine *N*-oxide (NMMO)[5] are either toxic, difficult to recycle or thermally unstable and therefore are not sustainable.

One way to introduce solubility as well as processability into cellulose is through chemical modification of its three hydroxyl groups (per anhydroglucose unit, AGU).[6] Such modifications, whether homogeneous or heterogeneous, has been investigated over the past years. Heterogeneous modifications are widely employed in industry since complete cellulose solubilization is not required, thus making the process much easier to apply.[7] On the other hand, the processes typically use over-stoichiometric amounts of reactants as well as acids or bases for the activation of the cellulose, thereby generating equal amount of waste. In addition, with this approach, control over the degree of substitution (DS) of the resulting modified cellulose is absent or difficult. In line with the twelve principles of green chemistry, which encourages the use of catalysis over stoichiometric reactants as well as avoidance of waste (*via* less derivatization),[8] homogeneous approaches are a better option offering the use of less reactants and resulting in a better control of the DS. However, overcoming the poor solubility of cellulose is a prerequisite. Thus, in order to keep the overall process of cellulose transformation sustainable, identifying a "Green solvent" for its solubilization and subsequent functionalization becomes a necessity.

*Cellulose and the search for Green solvents*

Identifying what criteria are necessary for a solvent to be considered "green" is usually not trivial. Notwithstanding, most researchers agree on the fact that a solvent with a very low vapour pressure, which is easily recyclable and re-usable, non-toxic, and if possible bio-derived falls in this category.[9] However, finding a solvent that meets all these criteria is not easy. Thus, a more general definition for a "green solvent" is one that leaves the least negative environmental and health footprint.[10] Therefore, simpler green metrics such as the ESH (environmental, health and safety) assessment method might come handy when making a decision between different solvents.[11] Hungerbühler *et al.* successfully applied this approach to 26 organic solvents in order to identify those with least environmental, safety and health concerns.[10] Even more so, there still appears a discrepancy between what researchers perceived as "green" solvents compared to what class of solvents they are working with. This was obvious when in 2011, Jessop conducted a survey among researches working in the field of green solvents. In the survey, the researchers were asked to choose what class of solvents would make the most impact on an overall decrease in environmental damages.[12] To his surprise, while the majority of the researchers chose mostly $CO_2$ and water as solvents, this contradicted the mostly employed ionic liquid solvents as reported in the Journal Green Chemistry in 2010. As many researchers mostly evaluate the success of a reaction based on yield, conversion and selectivity, a less "greener" solvent will easily be chosen if it meets these criteria compared to a more "greener" alternative. Hence, while a "greener" solvent might be interesting, there is the need to evaluate the entire process in order to avoid an "environmental burden shift" further down in the process cycle. Therefore, sustainability considerations are necessary as they take into consideration the entire process. To fully describe the sustainability of a process, a life cycle assessment (LCA) is advised, though it takes a considerable amount of time and the absence of data also makes its implementation challenging. However, by critically evaluating a given process, it is possible to identify better options (solvent, reactant and functionalization method), that will make the process more sustainable. The latter approach will be applied in the course of this review. Therefore, not only the "greenness" of the solvent will be considered; rather the entire process during cellulose transformation in order to identify more sustainable alternatives will be explored.

*Homogeneous Cellulose modification*

Many reports exist on homogeneous modification cellulose such as esterification/transesterification, alkoxylcarbonylation and etherification. However, sustainability assessment is usually not incorporated in most of these studies. Considering that the utilization of renewable resources is not enough to ensure sustainability,[13] we have taken a critical look into researches on the modification of cellulose in the line of sustainability. In this regard, selected examples of reports on cellulose modification will be discoursed in terms of the solvent used and its recycling/re-usability, the derivatization agent and its equivalents employed. In addition, the choice of chemical modification route employed is essential to ensure an overall sustainability of the process. In essence, in reviewing these reports, the principles of green chemistry as proposed by Anastas and Werner[8] will be highlighted and discussed. A general overview of reports on homogenous cellulose modification is shown in scheme 1.

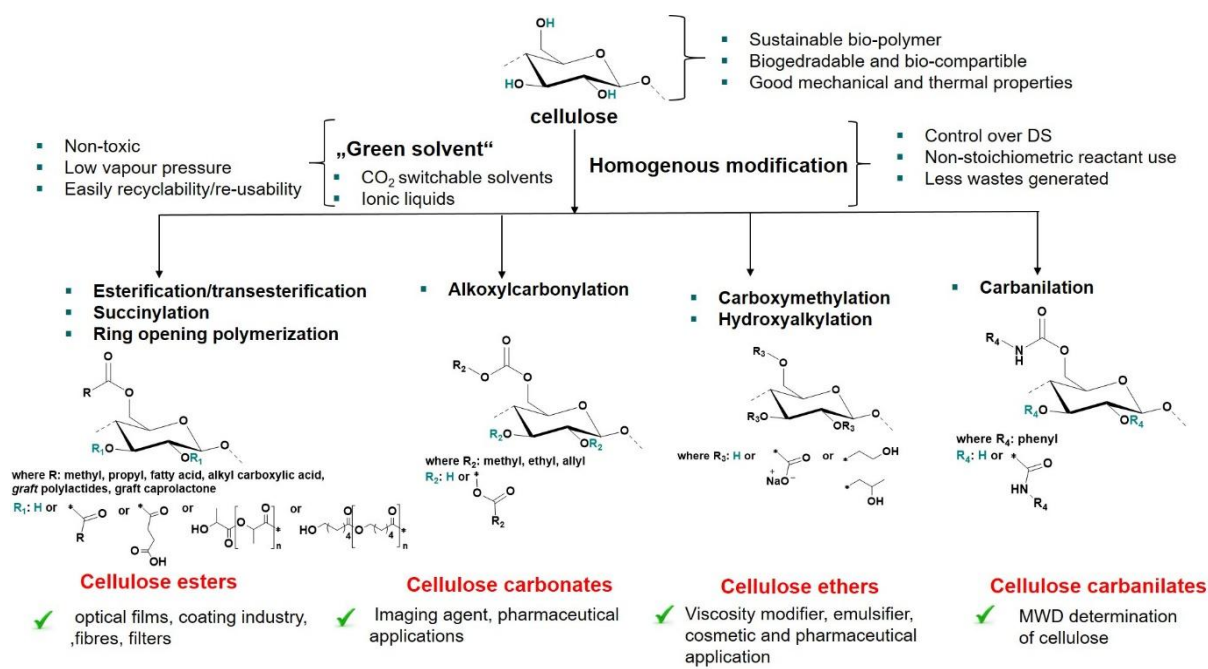

**Scheme 1**: *Cellulose derivatives through homogenous modification in ionic liquids or $CO_2$ switchable solvent system*

*Esterification and Transesterification*

Cellulose esters are among the most important and common forms of cellulose derivatives. They find applications in the coating industries and as optical films.[1] Typically, cellulose esters such as cellulose acetates are synthesized industrially *via* heterogeneous approaches using activated acids (acetic anhydride) in acetic acid as solvent.[7,14] This approach is considered more attractive for the industries compared to homogeneous routes, as it avoids the challenge of solubilizing the cellulose. However, this comes with an overall unsustainability due to the over-stoichiometric use of derivatization agent and also the use of sulphuric acid as catalyst to activate the cellulose hydroxyl groups, leading to generation of waste. While homogeneous acetylation of cellulose has been reported in classical cellulose solvents such as DMAC-LiCl,[3] and TBAF-DMSO,[4] their toxicity and difficult recyclability makes them non-sustainable. Therefore, the "greener" cellulose solvents such as ionic liquids and the more recent $CO_2$ switchable solvent system will be the focus of the present review. This class of solvents share the advantage of a low vapour pressure and the possibility to be recycled.

Ionic liquids are a class of solvent defined as molten salts with melting points below 100°C.[15] They are a broad class of solvents and typically consist of a relatively large acidic cation and a smaller basic anion. Graenacher had earlier reported the use of molten salts (alkyl pyrimidinium chlorides) for solubilizing cellulose.[16] However, since the melting point of such salts is above 100 °C, they are not classified as "ionic liquids" as defined in the present literature. Ionic liquids did not gain much interest as cellulose solvents until the report of Rogers *et al,* where high weight cellulose solubilization was demonstrated in the ionic liquid 1-*N*-butyl-3-methylimidazolium chloride ([$C_4$mim]$^+$Cl$^-$) without any derivatization.[17] Due to their very low vapour pressure and easy recyclability, ionic liquids have been promoted as "green solvents" for cellulose. The most common ionic liquids employed for cellulose solubilization consist of dialkyl imidazolium cations along with acetate or halogenide anions as shown in figure 1.

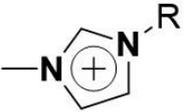

*Figure 1*: *Structures of frequently employed ionic liquids for cellulose solubilization*

Despite their wide success as cellulose solvents over such a short period of time, it is worth noting that ionic liquids are not without limitations.[18] Among these is their non-inertness (those with acetate anion such as [$C_2$mim]$^+$[OAc$^-$]), or corrosive and slight toxicity as in the case of [$C_4$mim]$^+$Cl$^-$. The non-inertness of ionic liquids makes their recovery difficult. Nonetheless, some reports have demonstrated the recovery and re-use of ionic liquids. BASF, which have a patent for [$C_2$mim]$^+$[OAc$^-$], reported the recovery of over 95% of the ionic liquid.[19] From a sustainability point of view, the large carbon footprint of their synthesis[20] can be justified if they can be recycled and re-used repeatedly. In addition, a careful choice of ionic liquids is necessary to avoid any possible side reaction that will make their recovery more problematic.

Ionic liquids have been employed for the homogeneous acetylation of cellulose. Guo *et al* investigated the acetylation of cellulose in the ionic liquid 1-*N*-allyl-3-methylimidazolium chloride ([Amim]$^+$Cl$^-$) using acetic anhydride.[21] A maximum DS of 2.74 was achieved using 5 eq. of the reactant/AGU of cellulose at 80°C within 23h. A good control of the DS was equally demonstrated by the variation of cellulose concentration, reactant equivalents and reaction temperature. Finally, the authors reported an easy recovery of the ionic liquid without any observable difference when applied for new cellulose solubilization and subsequent acetylation. While this report addressed some important sustainability concerns by the use of low vapour pressure solvents and a demonstration of their recovery and re-use, the use of activated acids such as acetic anhydride is not encouraged due to their instability. Furthermore, acetic acid as side product could promote hydrolysis of the cellulose backbone. Although the reaction was reported to occur in the absence of a catalyst, subsequent studies have shown that the presence of possible imidazole impurities in such ionic liquids could act as a catalyst. The purity of ionic liquids has been highlighted as one of its inherent challenges.[22] Reports have been shown of different batches of ionic liquids from the same manufacturer showing a different cellulose solubilization capability and reactivity behaviour.[23] In a similar approach, Barthel and Heinze reported acetylation of cellulose in the ionic liquids 1-*N*-butyl-3-methylimidazolium chloride ([$C_4$mim]$^+$Cl$^-$), 1-*N*-ethyl-3-methylimidazolium chloride ([$C_2$mim]$^+$Cl$^-$), 1-*N*-butyldimethylimidazolium chloride ([$C_4$dmim]$^+$Cl$^-$) and 1-*N*-allyl-2,3-dimethylimidazolium bromide ([Admim]$^+$Br$^-$).[24] Cellulose (microcrystalline cellulose, sulphite spruce pulp and cotton linters) was solubilized in these ionic liquids at 80°C for up to 12h. The reaction was performed using acetic anhydride or acetyl chloride in the presence or absence of pyridine as catalyst at 80°C. Cellulose solubilization was achieved within 2h at 80°C in [C$_4$mim]$^+$Cl$^-$. The authors reported a DS value of 3.0 within 30 min when acetyl chloride was used. In addition, a maximum DS value of 1.54 was reached when fatty acid chlorides such as lauroyl chloride were employed. The recovery of the ionic liquids was equally demonstrated and showed no loss in efficiency for subsequent cellulose solubilization. However, care must be taken to ensure complete water removal from the recovered ionic liquid, hence freeze-drying was proposed by the authors. Furthermore, the authors reported a degradation of the ionic liquids ([C$_2$mim]$^+$Cl$^-$, [Admim]$^+$Br$^-$,[C$_4$dmim]$^+$Cl$^-$) and the cellulose polymer when acetyl chloride was utilized. This undesirable observation might be due to the formation of HCl as side product when acetyl chloride is used. This not only reduces the recovery yield of the ionic liquids, but also the sustainability of the process.

In 2010, MarcFarlane *et al* demonstrated the concept of distillable ionic liquids for biopolymer processing.[25] Herein, tannin extraction was achieved from certain plant species using *N,N*-dimethylammonium-*N',N'*-dimethylcarbamate (DIMCARB) which was formed by the reaction of dimethylamine and CO$_2$. The adaption of this approach for cellulose solubilization was reported by Kilpeläinen *et al*.[26] In their work, they showed that a 1:1 molar ratio combination of an organic super base (1,1,3,3-tetramethylguadinine, TMG) with carboxylic acids such as formic, acetic or propionic acid results in an acid-base conjugate type ionic liquid capable of solubilizing cellulose. In the case of TMG in combination with acetic acid, the resulting [TMGH]$^+$[CO$_2$Et]$^-$ was capable of solubilizing 5 wt.% MCC (microcrystalline cellulose) within 10 min at 100°C. However, higher weight concentration (up to 10 wt.%) required much longer solubilization times (up to 20h) at the same temperature. More interesting in this report was the demonstration of the recycling of this ionic liquid with high purity and a recovery of over 99% by distillation between 100-200 °C and 1.0 mmHg pressure. As demonstrated, above a given temperature, the formed acid-base conjugate ionic liquid dissociates, allowing the separate species to be recovered as depicted in scheme 2.

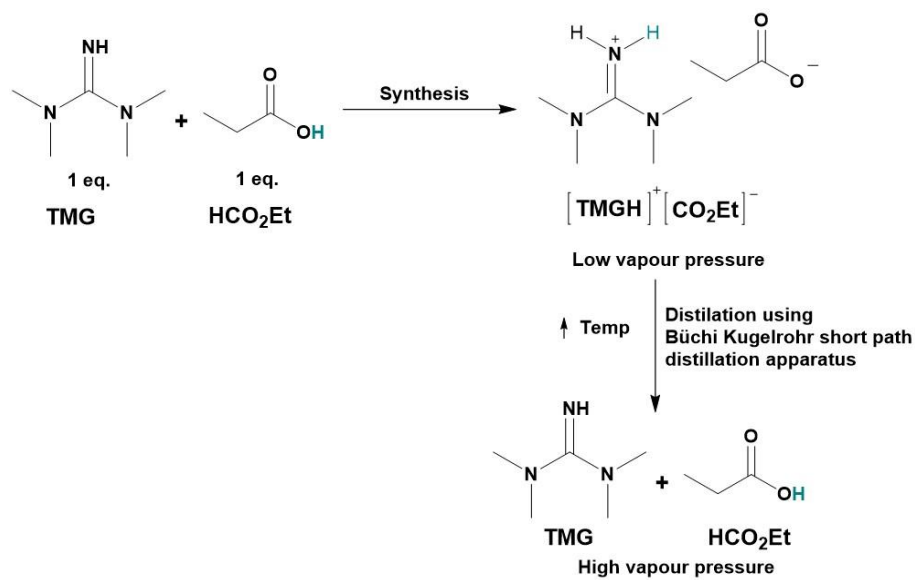

**Scheme 2**: Depiction of the concept of distillable ionic liquid from TMG and propanoic acid, adapted from Kilpeläinen et al.[26]

While this property is desirable for easy ionic liquid recovery, it also raises the question concerning the volatility of such a "ionic liquid" and its resulting constituents (high vapour pressure). This is because a low vapour pressure is considered as a key property for "green solvents", a property to which classical ionic liquids are ascribed.

The same authors subsequently reported the acetylation of cellulose (pre-hydrolysis kraft, PHK) in a similar acid-base conjugate ionic liquid.[27] In this case, the ionic liquid was formed by a 1:1 molar combination of the organic super base 1,5-diazabicyclo[4.3.0]non-5-ene (DBN) and acetic acid, resulting in the formation of [DBNH]$^+$[CO$_2$Et]$^-$. Acetylation was carried out after cellulose solubilization at 70°C for 0.5-1h using activated acids (acetic anhydride, propionic anhydride), vinyl esters (vinyl propionate, vinyl acetate) and isopropenyl acetate. When using acetic anhydride, acetic acid was formed as side product, thus requiring a slight excess of the base to trap this acid and prevent possible cellulose hydrolysis. Acetone and acetaldehydes were the observed side products when isopropenyl acetate and vinyl acetate or propionate were employed respectively. These side products could be removed easily, thereby keeping the equilibrium more on the product side. As reported by the authors, no extra catalyst was required compared to similar reports on acetylation of cellulose in ionic liquids requiring pyridine, K$_2$CO$_3$,[28] or NaOH[29] as catalyst. This was attributed to the excess of the base utilized in the ionic liquid which can act as a catalyst for the acetylation reaction. For the studies with acetic anhydride or isopropenyl acetate, DS of up to 3.0 were achieved. However, in the case of the vinyl esters, a maximum DS of 1.5 was reached and could not be improved by longer reaction time, higher temperature or excess of the reactant. Finally, the recovery of the ionic liquid was demonstrated (80% reported).[27] The authors observed the hydrolysis of DBN during the ionic liquid recovery. To minimize this hydrolysis, they proposed the use of n-butanol to trap remaining water molecules before distillation. Their work addresses sustainability in many aspects, from the choice of solvent, its recovery and mild acetylation conditions. However, it is worth pointing out that the use of activated acids such as acetic anhydride releases acetic acid that needs to be sequestered to avoid possible chain hydrolysis. The authors observed a decrease of the original weight of the cellulose pulp (up to 1/3 decreased after esterification) that could be due to the effect of the released acetic acid. Furthermore, the use of the highly reactive vinylic esters or isopropenyl acetate (expensive) could also lead to such degradation. In addition, these acetylating agents need to be synthesized before use, thereby increasing the carbon footprint of the entire process. Thus, improving the sustainability of this process could be achieved by a direct use of esters, that are mild and easier to handle.

It is important to note that reports of acetylation of cellulose in salt melts (eutectic mixture of KSCN, NaSCN and LiSCN·$H_2O$) have been reported.[30] However, in such solvents a very large excess of the derivatization agent (up to 50-100 equivalents/AGU) was required to achieve a DS of 2.4 after 3h at 130 °C. Such excess of reagents makes the process unsustainable and not practicable.

$CO_2$ switchable solvent systems are a recent class of cellulose solvents. Jessop *et al.* introduced the concept of utilizing $CO_2$ as a switch in the presence of an organic super base such as 1,8-diazabicyclo[5.4.0]undec-7-ene (DBU) to change the solvents polarity. However, the groups of Xie[31] and Jerome[32] simultaneously introduced the application of this concept for cellulose solubilization separately. A closer look into their reports, pointed out two classes of such $CO_2$ switchable cellulose solvents: derivative and non-derivative systems. In the derivative approach, as shown by Jerome *et al.*[32], cellulose could be solubilized by first transforming it into *in-situ* cellulose carbonate in the presence of a super base (examples include DBN, DBU, and TMG), which thus becomes soluble in DMSO. On the other hand,

Xie *et al.*[31] described the so-called non-derivative approach, where, instead of using cellulose directly, ethylene glycol was used as an alcohol source. In this case, the so-called $CO_2$ switchable solvent is prepared first by the reaction of the alcohol in the presence of a super base and DMSO. The formed solvent is then used for solubilizing cellulose. These two approaches are shown in scheme 3.

**Derivative approach**

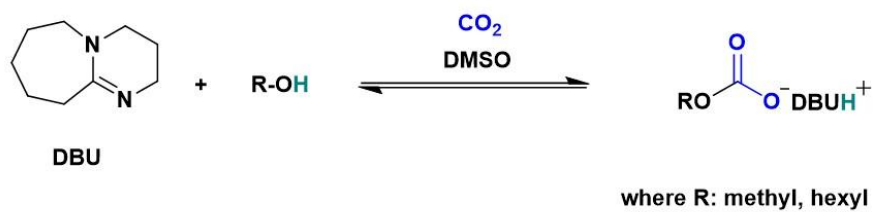

where R: methyl, hexyl

**Non-derivative approach**

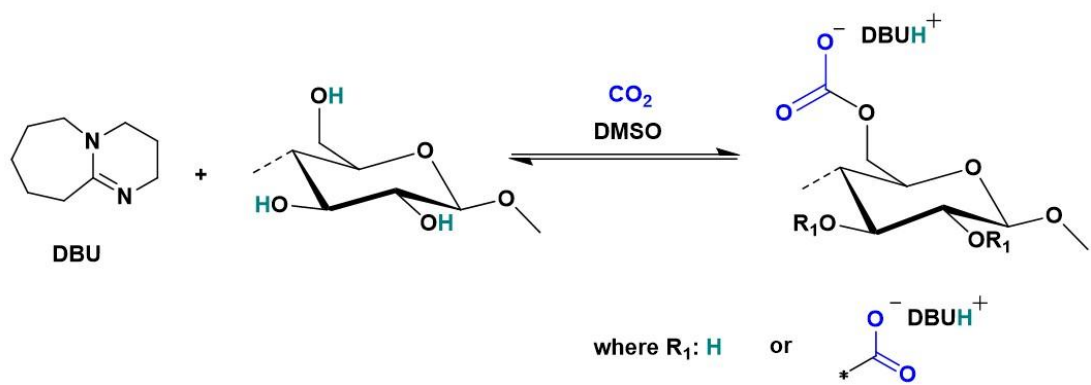

where $R_1$: H or [DBUH+ carbonate group]

**Scheme 3**: Derivative and non-derivative approach of $CO_2$ switchable solvent system, adapted from Xie et al[31] and Jerome et al.[32]

In both approaches, the solvents could be recovered by simply releasing the $CO_2$ pressure allowing the possible re-use of the solvent for new solubilization. Furthermore, the relatively cheap cost of the entire solvent system compared to traditional ionic liquids coupled with their lower toxicity compared to DMAC-LiCl makes these solvents more sustainable. In our opinion, the so-called non-derivative approach employing the use of simpler alcohols is not necessary. The extra steps involved through this approach simply generates more wastes as side reactions becomes unavoidable. Recently, we demonstrated a more detailed understanding of the non-derivative $CO_2$ solvent system using DBU as super base. Upon optimization, complete cellulose solubilization (up to 10 wt.%) was achieved at 30 °C within 10-15 min under low $CO_2$ pressure (2-5 bar).[33] Furthermore, the presence of the *in-situ* formed cellulose carbonate was un-ambiguously proven by trapping it with electrophiles.[33] The fast solubilization time and milder conditions as demonstrated for this class of cellulose solvent, in our opinion could pave a way for their wide spread utilization for various homogenous modification of cellulose.

The utilization of this solvent system for homogeneous cellulose modification has been reported. In this regard, the groups of Xie and Liu employed the non-derivatizing $CO_2$ switchable solvent system for the acylation of cellulose under mild conditions.[34] High DS values of cellulose acetates could be reached using acetic anhydride. Furthermore, other types of cellulose esters such as cellulose butyrate and propionate could be equally synthesized. Interestingly and opposed to other solvent systems such as ionic liquids or DMAC-LiCl, where a catalyst was necessary for such modifications, the authors showed that the acylation in this solvent system required no extra catalyst. This was attributed to the dual role of DBU, which is part of the solvent system and equally acts as an organocatalyst for the acylation step. It is also important to mention the demonstration of recovery of the solvent system, which is an important consideration for sustainability. However, a slight drawback this procedure was the observed side reaction between methanol and the acetic anhydride leading to the formation of methyl acetate. Even though this side product could be removed, it undermines the entire process since a higher equivalent of both methanol and acetic anhydride will be required to account for this loss. Just as mentioned earlier, the use of such a non-derivative $CO_2$ solvent system is not encouraged. On the other hand, as the derivative approach directly utilizes the cellulose, such undesirable side reaction can be completely avoided, thus more sustainable. The same authors also reported a fast and mild esterification of cellulose pulp (cotton) using acetic anhydride.[35] This time, the derivative $CO_2$ switchable solvent approach was utilized.[35] As in previous examples, no need for an extra catalyst was required. Furthermore, they showed a good control of the DS for the obtained cellulose esters by simply varying reaction parameters such as reaction time, temperature and equivalents of the acetylating agent. More so, good recovery of the solvent system (DMSO and DBU) was demonstrated.

In the previous reviewed reports, we noticed that activated acids such as acetyl chlorides, acetic anhydride or vinyl acetates are frequently employed for acetylation of cellulose. Some of these reactants are toxic (acetyl chloride), but they are also unstable and difficult to handle compared to esters. Furthermore, as can be seen in some of these reports, the associated side products such as HCl or acetic acid could lead to polymer degradation. Therefore, in line with the principles of green chemistry, the use of benign reactants is encouraged. In this regard, the groups of Meier and Barner-Kowollik employed the use of methyl esters for the synthesis of various cellulose esters such as cellulose butyrate and cellulose benzoates using catalytic amounts of TBD.[36] Although the reaction was performed in a traditional ionic liquid ([$C_4$mim]$^+$Cl$^-$), the authors demonstrated its recovery and re-use. Furthermore, the sustainability of this process could be improved by directly utilizing plant oils instead of fatty acid methyl esters (FAMEs). This approach will avoid the associated derivative steps along with the waste involved in synthesizing FAMEs from plant oils. In this light, we demonstrated a direct transesterification of cellulose using plant oils (high oleic sunflower oil) in the derivative DBU-$CO_2$ switchable solvent system.[37] A DS of 1.59 was achieved after 24h and 115°C when 3.0 equivalents of plant oil per AGU of cellulose were utilized. In addition, no additional catalyst was required as the DBU utilized in the solvent system also acted as an organocatalyst for the transesterification process. Finally, this approach was successfully transferred to different cellulose types such as MCC, filter paper and pulp.

*Functional cellulose esters*

Ring opening polymerization (ROP) or "grafting from" is one approach for obtaining functional cellulose esters. This can be achieved by using bio-derived and sustainable reactants such as lactides (lactone cyclic di-esters) or ε-caprolactones (cyclic esters). Lactides can be obtain from lactic acid (2-hydroxy propanoic acid) that are derived from sour milk or through fermentation of carbohydrates (starch or simple sugars),[38] whereas ε-caprolactones is derived from caproic acid (hexanoic acid). The groups of Xie and Liu employed the DBU-$CO_2$ switchable solvent system for grafting of L-lactide from cellulose *via* ROP.[39] In this report, no additional catalyst was required as the DBU again played a dual role: as part of the solvent system and equally as an organocatalyst during the ROP. A higher grafting density was obtained compared to previous reports in different solvents such as DMAC-LiCl[40] or ([Amim]$^+$Cl$^-$).[41] Furthermore, their approach is more sustainable than that of previous reports in ([Amim]$^+$Cl$^-$), where tin(II) 2-ethylhexanoate (Sn(oct)$_2$) was employed as catalyst with a lower grafting density reported.[41] The obtained cellulose-*g*-poly(L-lactide) reported by Liu and co-workers could be solubilized in water and also showed tunable $T_g$ properties. In a similar fashion, the same authors reported on the ROP of ε-caprolactones from cellulose in the DBU-$CO_2$ switchable solvent system.[42] As in the previous report, no need for any extra catalyst was required as DBU also acted as an organocatalyst. In addition, the authors reported a high value $DS_{PCL}$ (2.83) and grafting efficiency (74.1%) compared to similar experiments in ionic liquids.[43] Furthermore, the synthesized cellulose-*g*-PCL (cellulose-graft-poly(ε-caprolactone) showed $T_g$ values between -35 °C to -53 °C, thereby addressing the non-processability challenge of native cellulose. However, in these studies, the authors did not report any study on the recovery of the solvent system (DBU and DMSO), which is an important consideration from a sustainable point of view

Succinylation of cellulose is another approach for obtaining functional cellulose materials bearing a pendant carboxylic acid group. In this case, succinic anhydride (which is bio-derived from sugars)[44] can be utilized. Many studies have reported the succinylation of cellulose in various solvents such as ionic liquids ($[C_4mim]^+Cl^-$),[45] DMAC-LiCl,[46] TBAF-DMSO[47] and tetrabutylammonium acetate-dimethyl sulfoxide (TBAA-DMSO).[48] In most of these reports, a catalyst such as 4-(dimethylamino) pyridine (DMAP)[45] or triethylamine[46] was required alongside high temperatures (100 °C) to achieve a high DS (up to 2.34). As previously mentioned, except for ionic liquids, these solvents are toxic, not easily recycled and hence not sustainable. Also absent in these studies was the recovery of the solvents, which is an important consideration for sustainability. Hence, the group of Meier and Cramail demonstrated a more sustainable approach for the succinylation of cellulose in the $CO_2$-DBU switchable solvent system.[49] The easy solubilization step (30 min at 50 °C), mild reaction conditions (room temperature, 30 min), low succinic anhydride equivalents (4.5 eq. per AGU of cellulose), coupled with the absence of an additional catalyst and the demonstration of the solvent (DBU and DMSO) recyclability makes this approach more sustainable. Interestingly, a high DS of 2.6 was achieved in this report, despite the milder reaction conditions employed without additional catalyst. Thus, this report clearly shows that sustainability principles can be fully applied in the course of renewable resource utilization.

*Alkoxylcarbonylation*

Cellulose carbonates are a very important class of material with potential application as imaging agents or for drug delivery.[50] However, early reports on the synthesis of cellulose carbonates employed toxic chloroformates. In addition, these chlorofomates are usually derived from phosgene, which is a toxic reagent and not stable. Thus, from a sustainable point of view, their use is not encouraged. In this regard, King *et al*, reported on using dialkylcarbonates such as diethyl carbonate (DEC) and dimethyl carbonate (DMC).[51] Generally, various dialkyl carbonates can be obtained from the reaction between DMC and alcohols in the presence of catalytic amounts of TBD (1,5,7-triazabicyclo[4.4.0]dec-5-ene).[52] At the same time, DMC can be obtained from the reaction between methanol and $CO_2$. In their report, King *et al*, employed the two ionic liquids 1-*N*-ethyl-3-methyl imidazolium acetate ($[C_2mim]^+[OAc]^-$) and trioctylphosphonium acetate ($[P_{8881}]^+[OAc]^-$). As with most reports utilizing ionic liquids, DMSO was added as a co-solvent to reduce the viscosity of the resulting solubilized cellulose. However, they noticed a reaction of the acetate anion of the ionic liquid with the DMC leading to the formation of methyl acetate, thus making the ionic liquids recovery very challenging. However, by using a milder recovery procedure, they were able to minimize this side reaction. Hence, while this study demonstrated the use of a more sustainable reagent for the synthesis of cellulose carbonate, it also brings to light the challenge faced by ionic liquids due to their non-inertness in some cellulose modification reactions. Hence, it is necessary to verify the inertness of the ionic liquid for a given chemical transformation.

As pointed out in the previous work by King *et al*.[51], ionic liquids containing the acetate anion are more prone to contamination in the course of most reactions. One way to avoid this is to use their chloride counterparts such as 1-*N*-butyl-3-methyl imidazolium chloride ($[C_4mim]^+Cl^-$) or 1-*N*-allyl-3-methyl imidazolium chloride ($[Amim]^+Cl^-$). This has been demonstrated by Söyler and Meier.[53] In this case, diallyl carbonate was employed for the modification of cellulose in ($[C_4mim]^+Cl^-$) in the presence of DMSO as co-solvent (10 w/w.% DMSO). Diallyl carbonate was synthesized from dimethyl carbonate and allyl alcohol in a benign fashion in one-step. The obtained allyl-functional cellulose carbonate was obtained with a DS of 1.3. Utilizing the allylic functionality, a post modification *via* thiol-ene was achieved.[53]

Interestingly, no reaction between the ionic liquid and DAC was observed, thus leading to a facile recovery of the ionic liquid, which could be re-used for new modifications. This report demonstrated the possibility to apply sustainability in the entire transformation of cellulose. In addition, it is important to note that for an ionic liquid to be considered sustainable, its inertness in the reaction should be ensured in order to enable an easy recovery. However, it is worth mentioning that ([$C_4$mim]$^+$Cl$^-$) is slightly acidic, corrosive and shows some toxicity. Therefore, its recovery after usage is strongly advised to avoid a release to the environment.

*Etherification*

Carboxyl methylcellulose (CMC) is among the commercially most used derivatives of cellulose ethers and finds application as viscosity modifiers and emulsion stabilizers.[54] Heinze *et al* demonstrated carboxymethylation of cellulose in the ionic liquids 1-*N*-butyl-3-methyl imidazolium chloride ([$C_4$mim]$^+$Cl$^-$) in the presence of DMSO as co-solvent.[55] A DS of 0.43 was reached when 1.0 equivalent of the reactant per AGU was used. Higher equivalents did not improve the DS value. The authors demonstrated the recovery and re-used of the ionic liquid. However, it is important to note that the use of sodium chloroacetate for the carboxymethylation is not sustainable. Furthermore, the use of NaOH as catalyst in this reaction required an acidic work-up for neutralization, which further generates waste and further reduces the sustainability of the process.

Quaternary ammonium electrolytes (QAEs) in the presence of molecular solvents such as DMSO have been reported for cellulose solubilization.[56] However, as mentioned earlier and using TBAF/DMSO as an example, the majority are considered toxic due to the presence of the halogen counter-ion. However, Heinze and his group reported a less toxic variant where the halogens are replaced by a carboxylate anion (e.g. formate).[57] In this case, the QAEs such as [$CH_3N(CH_2CH_3)_3$]$^+$[HCOO]$^-$ were obtained by quaternization of tertiary amines using benign dialkyl carbonate reagents. The efficiency of this solvent system was demonstrated for the homogeneous carboxymethylation of cellulose using sodium chloroacetate. A water soluble CMC with a DS of 1.55 was obtained.[57] Furthermore, an attempt to recover the solvent was reported. However, only the cation could be recovered, as the carboxylate anion was exchanged by the chloride ion from NaCl formed as side product, thus preventing full recovery. As in the previous examples, the use of sodium chloroacetate is not sustainable, just as the use of inorganic sodium salts, which generated salt wastes. Thus, the replacement of sodium chloroacetate as well an organic basic catalyst that can be recovered will improve the sustainability of this process.

Hydroxyalkylation is another approach to obtain cellulose ethers. Such cellulose ethers find application in the paint and pharmaceutical industries as well as in the material construction industries.[54] By slowing down the cement hydration, they lead to improved mechanical properties.[58] Heinze *et al* reported the hydroxyalkylation of cellulose using ethylene and propylene oxides.[59] The reaction was carried out in the ionic liquid [$C_2$imim]$^+$[OAc]$^-$. In this case, the basic acetate anion was able to catalyze the reaction, thereby preventing the use of any extra inorganic base catalyst and a rather high DS of 2.79 was reached. However, since ethylene or propylene oxides (oxiranes) are toxic, replacing them in the near future will improve the sustainability of the process.

*Carbanilation of cellulose*

Cellulose characterization *via* SEC in order to obtain its molecular weight distribution (MWD) is not possible due to its insolubility in common SEC solvent.[60] Therefore, carbanilation of cellulose offers a possibility to obtain soluble polymers for SEC analysis. Typically, the reaction is quantitative and a DS of 3.0 can be achieved easily, as shown by Heinze *et al*, in the ionic liquid ([$C_4$imim]$^+$[Cl]$^-$).[24] However, this process employs the use of toxic isocyanates (phenyl isocyanates), which are usually derived from toxic phosgene. Even though a phosgene-free route for the synthesis of isocyanates has been reported by Knölker *et al*,[61] the toxicity of the isocyanates is still worth being taken into consideration.

**Perspective: the future of sustainable homogeneous cellulose modification**

Recently, and celebrating the 20$^{th}$ anniversary since the introduction of the twelve principles of green chemistry, Anastas *et al* published a detailed review highlighting what progresses have been made in the field of green chemistry.[62] As rightly pointed out in their work, researchers working in the field of green chemistry are advised to avoid an "environmental burden shifting". To address this, the twelve principles are not to be considered in isolation, but rather a careful evaluation of the entire transformation process is necessary. In this regard, while the use of cellulose as a raw material meets the 7$^{th}$ principle of green chemistry, which aims at promoting the use of renewable resources, avoiding an "environmental burden shifting" is important.

Since "design" can be seen as the centre around which the twelve principles of green chemistry rotate, achieving sustainable cellulose modification can be realized by a careful design. Therefore, we propose four important questions that are worth being answered before cellulose modification in order to ensure sustainability. In the first question, the type of solvent to be used need to be addressed. As special solvents are required for cellulose solubilization, important considerations such as toxicity, recyclable/re-usability, bio-derived, required temperature, and solubilization time needs to be taken into account. Through this way, comparison among various solvents can be made in order to choose the "greenest" option. Secondly, the choice of the derivatization agent (reactant) needs to be considered. In this regard, evaluation on their toxicity, source, equivalents required, possible side reactions, easy separation from desired product as well as their stability are very important. Also important to answer as the third question is the choice of the functionalization method. Herein, considerations on the use of catalyst over stoichiometry, atom efficiency, yield, absence of protection groups or pre-derivatization step, milder reaction conditions and easy recovery of product are necessary. Finally, as a fourth question, the obtained modified cellulose is equally important. In this case, the issue of toxicity, biodegradability (to what type of products) and stability are important to consider. To ensure sustainability therefore during cellulose transformation, these proposed group of questions needs to be considered together and not in isolation. Thus, through design, the entire process from "cradle to grave" can be put into perspective, thereby achieving sustainable cellulose-based materials. This design approach during cellulose modification is summarized in Scheme 4. In addition, we have highlighted the principles of green chemistry that are being addressed by employing these suggestions. In summary, careful consideration of the entire transformation process of cellulose is necessary to ensure sustainability.

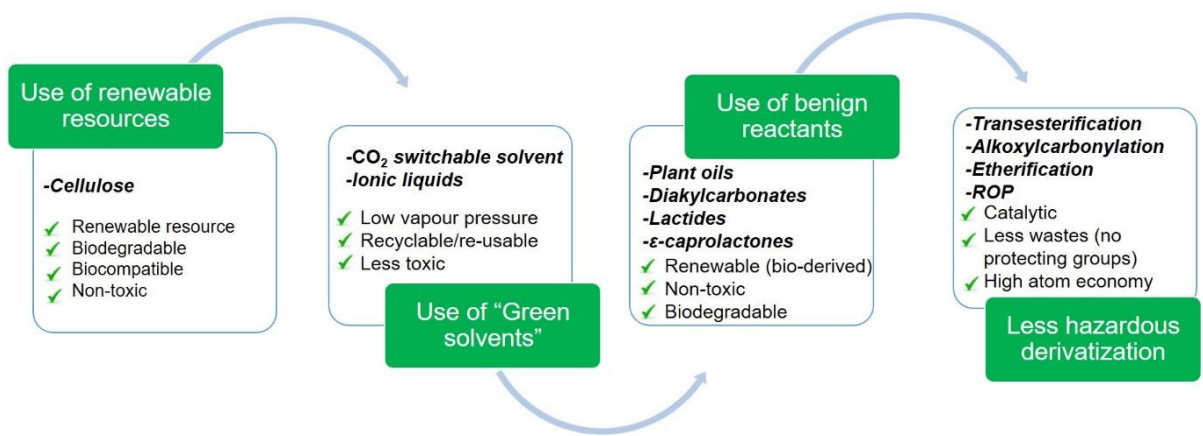

***Scheme 4***: *Design approach towards ensuring sustainability during cellulose modification*

**Conclusions**

As the most abundant bio-based polymer, cellulose holds a great potential as a viable alternative to replace the non-sustainable fossil derived resources. However, despite its renewability, there is a need for sustainability issues to be taken into consideration during their chemical transformation in order to produce materials of interest. In line with the twelve principles of green chemistry, the design of cellulose-based materials will address environmental challenges if its transformation avoids the use of toxic reactants or solvents as well as harsh reaction conditions. Thus, at the beginning of any intended chemical transformation of cellulose, the issue of sustainability is worth considering. While a quantitative approach such as LCA are difficult to apply due to their time demand and a lack of sufficient data, simpler green metrics such as E-factor, EHS (environmental, health and safety) and atom economy can be employed to choose between solvents, reactants and functionalization approaches during cellulose transformation. As one of the primary goal of modification of cellulose is to make processable materials that will replace their non-sustainable fossil-based counterpart, therefore ensuring sustainability in their entire transformation is of utmost importance and necessity.